\documentclass[fleqn,12pt,twoside]{article}
\usepackage{espcrc1}
\usepackage{graphicx}

\newcommand{\AmS}{{\protect\the\textfont2
  A\kern-.1667em\lower.5ex\hbox{M}\kern-.125emS}}

\title{Hybrid potential/R-matrix models for the $^{12}$C~+~$\alpha$ system}

\author{Jean-Marc Sparenberg\address{TRIUMF Theory Group, 4004 Wesbrook Mall,
Vancouver, BC, Canada V6T 2A3}}
       
\begin{document}

\maketitle

\begin{abstract}
I present ongoing developments on hybrid models which combine the potential
model, for the description of the background, with the R-matrix model, for the
description of discrete states.
Physical parameters of discrete states are converted into formal R-matrix
parameters, taking into account the potential background. 
The nuclear part of this potential is constructed by a new variant of an
inversion method based on supersymmetric quantum mechanics.
The method is illustrated on $^{12}$C~+~$\alpha$ $S$-wave elastic phase shifts.
\end{abstract}

\section{MOTIVATIONS}
In nuclear astrophysics, the key issue is the determination of cross
sections at very low energies, representative of stellar conditions.
Because of the Coulomb repulsion between nuclei, these cross sections
are sometimes minuscule, in particular for non-resonant processes,
and hence very hard to measure.
Nuclear theory has thus an important role to play for the extrapolation of
measured data down to astrophysical energies.
A reliable extrapolation requires (i) a precise fit of all available data,
(ii) a good account for both resonant and non-resonant processes,
(iii) a small number of parameters to limit the extrapolation uncertainty.
The R-matrix phenomenological model \cite{descouvemont:01}
is the only one to date which is able to precisely fit experimental data;
however, whereas its description of discrete (resonant) states is satisfactory,
its description of the non-resonant background,
based on phenomenological high-energy wide states,
requires a lot of additional parameters and lacks physical motivation.

I demonstrate here the interest of the potential model for a better description
of the background,
complemented by an R-matrix phenomenological description of discrete states.
Such hybrid models have already been used in previous works
\cite{westin:71,langanke:85}
but the model developed here presents two new features.
First (section 2), discrete states are characterized by physical parameters (energy and width) rather than by formal R-matrix parameters,
a procedure recently advocated for in the traditional R-matrix method
\cite{angulo:00,brune:02}.
Second (section 3),
the potential model is constructed by inversion of scattering data,
which leads to a better-quality fit than phenomenological potentials like
the Woods-Saxon one.

As an example, I apply the method to the $S$-wave phase shifts
deduced \cite{buchmann:03}
from a recent high-precision measurement of the $^{12}$C~+~$\alpha$
elastic scattering \cite{tischhauser:02}.
This analysis is preliminary to the analysis of the $P$ and $D$ waves for
this system, which dominate the capture reaction at astrophysical energies.
For these waves, the present method could lead to a better description
of the interference effects between the background and the subthreshold states,
which are known to strongly influence the reaction rate at astrophysical
energies.
The $2^+$ subthreshold state has recently been studied in the
framework of the inversion potential model \cite{sparenberg:04a} but a
generalization to a hybrid model is necessary to study the $1^-$ subthreshold
state and to get a precise fit of all available data.

\section{HYBRID MODEL WITH PHYSICAL PARAMETERS}

\begin{figure}[t]
\begin{minipage}[t]{80mm}
\framebox[79mm]{\includegraphics[scale=0.45]{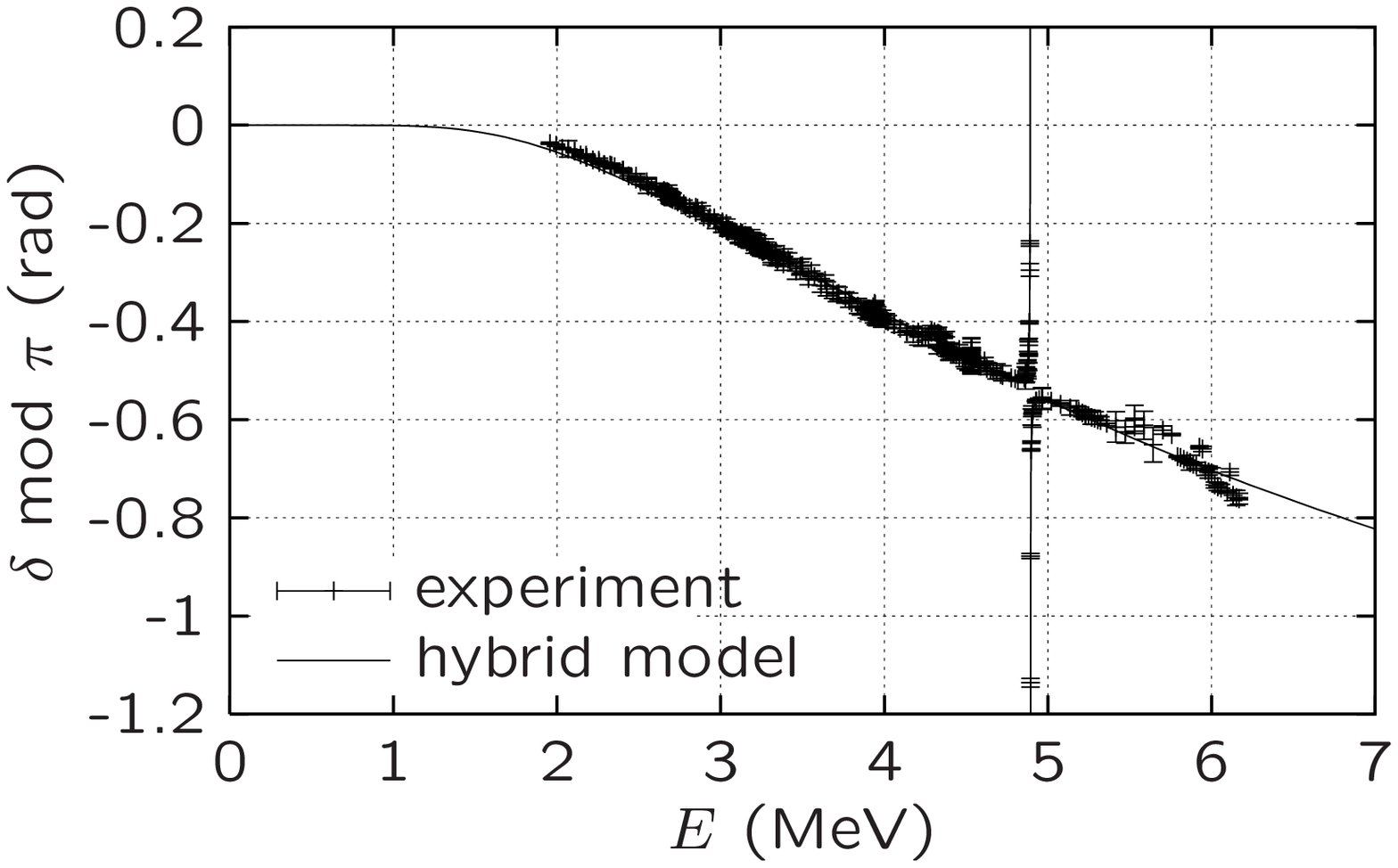}}
\caption{Fit of experimental $^{12}$C~+~$\alpha$ $S$-wave
elastic-scattering phase shifts \cite{buchmann:03,tischhauser:02}
by a hybrid model consisting of a soft-sphere potential and one R-matrix pole.}
\label{d0hybbw}
\end{minipage}
\hspace{\fill}
\begin{minipage}[t]{75mm}
\framebox[79mm]{\includegraphics[scale=0.45]{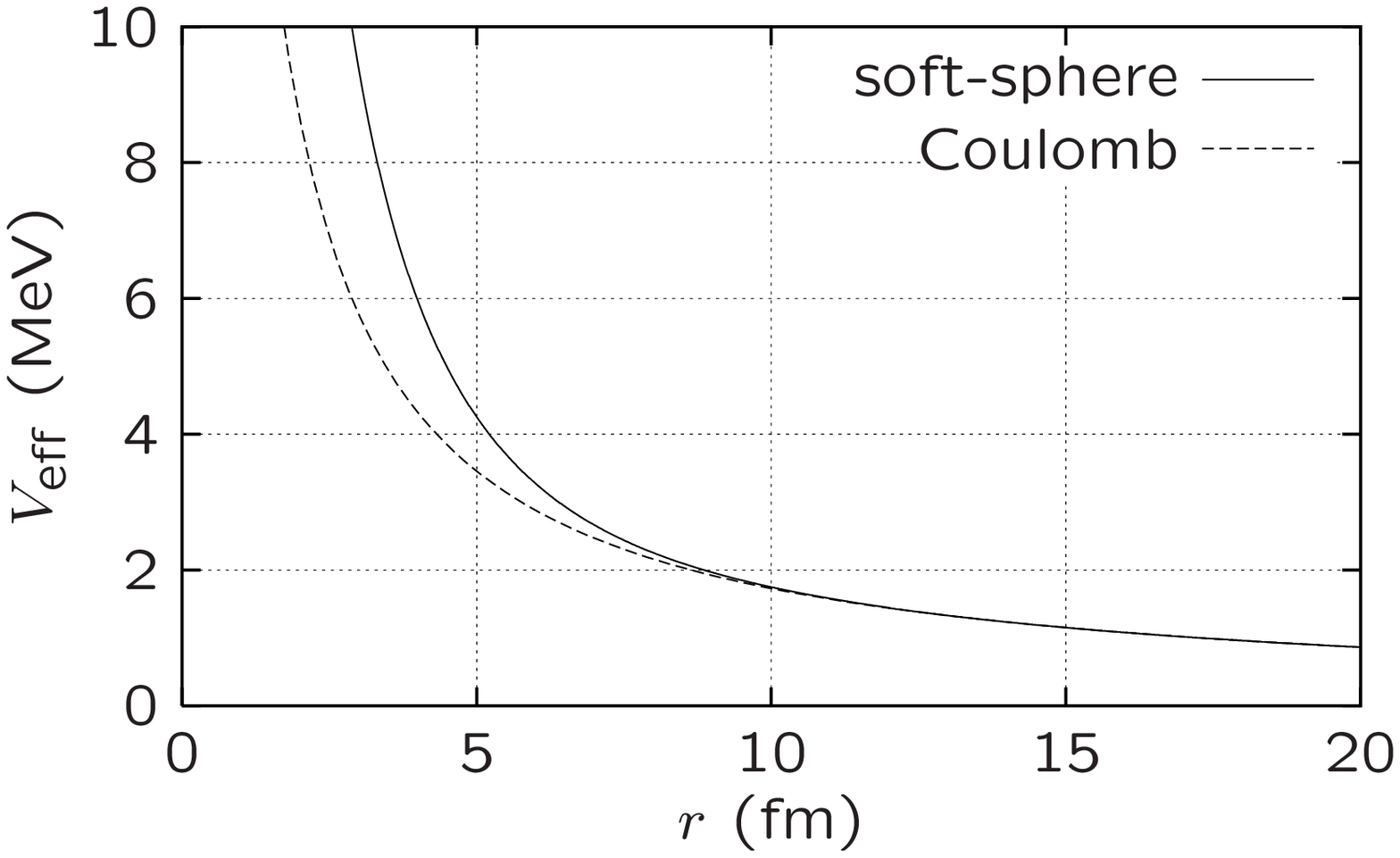}}
\caption{Soft-sphere $^{12}$C~+~$\alpha$
effective potential ($\ell=0$) used for the fit of Fig.\ \ref{d0hybbw},
compared with the pure Coulomb potential.}
\label{p0hybbw}
\end{minipage}
\end{figure}

I use the hybrid potential/R-matrix model presented in refs.\
\cite{westin:71,langanke:85}.
The general principle is to calculate, for a given partial wave,
the R-matrix corresponding to the potential model,
$R_\mathrm{pot}(E)$, and to add to it phenomenological poles
corresponding to discrete states.
This leads to the total R-matrix
\begin{eqnarray}
R(E) = R_\mathrm{pot}(E) + R_\mathrm{res}(E)
= R_\mathrm{pot}(E)+\sum_{\lambda=1}^L\frac{\gamma_\lambda^2}{E_\lambda-E},
\label{Rhyb}
\end{eqnarray}
where $E$ is the center-of-mass energy and only one channel is considered for
simplicity (this is sufficient for $^{12}$C~+~$\alpha$ elastic scattering).
Discrete states are characterized by their formal energy $E_\lambda$
and reduced width $\gamma_\lambda^2$.
The main advantage of the above formula is its simplicity,
which also implies a very convenient use in numerical calculations.
Its drawback is that the formal parameters cannot be directly interpreted as
physical state properties: their value depend on the R-matrix radius and
boundary parameter $B$.
This creates strong parameter correlations in R-matrix numerical fits
where the R-matrix radius itself can be varied.
It is thus preferable to use physical parameters $\widetilde{E}_\lambda$,
$\widetilde{\gamma}_\lambda^2$ as fitting parameters and to
convert them into formal parameters for the actual calculation of the fitted
quantity.
In the usual R-matrix formalism,
the scattering (S) matrix is factorized into a hard-sphere
(hs) factor and an R-matrix factor,
the behavior of the latter around resonances defining physical parameters:
\begin{eqnarray}
U(E)=e^{2\imath\delta(E)}=U_\mathrm{hs}(E) \times \frac{1-[L^*(E)-B]R(E)}
{1-[L(E)-B]R(E)},
\quad L(E)=S(E)+\imath P(E),
\label{Srmat}
\end{eqnarray}
where $\delta$ is the phase shift,
and $S$ and $P$ are the shift and penetration factors,
respectively~\cite{descouvemont:01}.
Here, we want to define physical parameters with respect to the potential
S matrix rather than to the hard-sphere one.
By inserting Eq.\ \ref{Rhyb} into Eq.\ \ref{Srmat},
one gets a different factorization for the S-matrix:
\begin{eqnarray}
U(E)=U_\mathrm{pot}(E) \times
\frac{1-\left\{[L^*(E)-B]^{-1}-R_\mathrm{pot}(E)\right\}^{-1}R_\mathrm{res}(E)}
{1-\left\{[L(E)-B]^{-1}-R_\mathrm{pot}(E)\right\}^{-1}R_\mathrm{res}(E)}.
\label{Shyb}
\end{eqnarray}
Comparison of the last factors of Eqs.\ \ref{Srmat} and \ref{Shyb} shows
that existing R-matrix methods transforming physical parameters into
formal ones
can directly be applied to the hybrid case, provided $[L(E)-B]$ is replaced by
$\left\{[L(E)-B]^{-1}-R_\mathrm{pot}(E)\right\}^{-1}$.
In refs.\ \cite{angulo:00,brune:02}, such methods are proposed that take into
account the interference of the different R-matrix poles with each other.
Although these methods could be generalized to the hybrid case,
the interference effects are much less important here since no wide
pole is used to describe the background.
Up to now, I have found sufficient to use the one-pole approximation for which,
for instance, the physical to formal energy conversion formula reads
\begin{eqnarray}
E_\lambda = \widetilde{E}_\lambda + \gamma_\lambda^2 \times
\frac{S(\widetilde{E}_\lambda)-B-R_\mathrm{pot}(\widetilde{E}_\lambda)
|L(\widetilde{E}_\lambda)-B|^2}
{1-2 R_\mathrm{pot}(\widetilde{E}_\lambda)[S(\widetilde{E}_\lambda)-B]
+ R_\mathrm{pot}^2(\widetilde{E}_\lambda) |L(\widetilde{E}_\lambda)-B|^2}.
\end{eqnarray}
The conversion formula for the width is too long to be given here.
These formulas are used for the $^{12}$C~+~$\alpha$ one-pole example of
Fig.\ \ref{d0hybbw},
for which I have checked that the best-fit physical parameters of the resonance
($\widetilde{E}_1=4.89271$ MeV,
$\widetilde{\Gamma}_1=2P(\widetilde{E}_1)\widetilde{\gamma}_1^2=1.32$ keV)
are practically independent of the R-matrix radius
(above 12 fm, see next section)
and hence very convenient fitting parameters,
contrary to formal parameters.

\section{POTENTIAL CONSTRUCTION BY INVERSION}

An inversion technique based on supersymmetric quantum mechanics has been
proposed in ref.\ \cite{sparenberg:97a} and has been successfully applied to
the $^{12}$C~+~$\alpha$ system in ref.\ \cite{sparenberg:04a}.
For a given partial wave,
this method allows the deductive construction of a potential reproducing
elastic-scattering phase shifts.
This method is based on a rational expansion
in the complex wave-number $k$-plane of the partial-wave S matrix
\begin{eqnarray}
U_\mathrm{pot}(E=k^2) = U_0(k)
\prod_{m=1}^M \frac{\kappa_m+k}{\kappa_m-k},
\end{eqnarray}
where the poles $\kappa_m$ are fitted to the phase shifts.
They can be determined either by resolution
of the linear system of equations giving the coefficients of the polynomials
appearing in the S matrix \cite{sparenberg:97a}
or with a general minimization procedure \cite{samsonov:02}.
This last method, though less deductive,
allows one to easily constrain the poles to stay on the imaginary axis,
which avoids oscillations in the obtained potential.
This effective potential is constructed iteratively by algebraic
(supersymmetric) transformations of the radial Schr\"odinger equation
$V_0(r) \rightarrow V_1(r) \rightarrow \dots \rightarrow V_M(r)$,
with a number of transformations equal to the number $M$ of poles of the
S matrix.

The potential $V_0$ and its corresponding S matrix $U_0$ are references which
can be chosen arbitrarily.
For a system without Coulomb interaction, one generally chooses $V_0=0$;
the supersymmetric transformations can then be performed analytically and the
final potential has a compact expression in terms of exponential functions.
For a system with Coulomb interaction, one generally chooses $V_0$ as the pure
Coulomb potential.
However, this has the drawback that individual transformations then
create long-range $r^{-2}$ terms in the potential.
These can be approximately canceled by successive transformations,
provided most S-matrix poles lie roughly on a circle around the origin
of the complex plane,
but this requires a rather large number of poles and generally creates small
oscillations in the potential \cite{sparenberg:97a}.
For instance, to perfectly reproduce the $^{12}$C~+~$\alpha$
$S$-wave phase shits represented in Fig.\ \ref{d0hybbw},
10 poles are needed for a pure Coulomb $V_0$.
This makes the construction of the potential rather heavy from the numerical
point of view, which is not very suitable for applications;
moreover, the potential presents slight oscillations.
An alternative method used here is to construct by supersymmetric
transformations the nuclear part of the potential 
instead of the total effective potential.
One starts from $V_0=0$, constructs the nuclear potential $V_M$,
adds Coulomb and centrifugal terms to it and only then calculates (nuclear)
phase shifts.
The fit can only be made by automatic minimization procedure in this case.
This method presents several advantages:
the nuclear potential is short-ranged, it has a compact analytical expression
and the number of poles is small, which facilitates numerical fits.
For instance, the fit presented in Fig.\ \ref{d0hybbw} has been obtained with
only two poles at wave number $\kappa_1=\kappa_2=0.4463 \imath$ fm$^{-1}$.
No significantly better fit could be obtained without introducing small
oscillations in the potential.

The corresponding effective potential is represented in Fig.\ \ref{p0hybbw},
together with the pure Coulomb potential.
$V_2$ is a purely repulsive {\em soft-sphere} potential 
(no bound or resonant state) becoming negligible above 12 fm.
This large range is rather surprising:
in usual R-matrix calculations \cite{tischhauser:02,sparenberg:04b},
the range of the nuclear interaction is generally
estimated between 5.5 and 6.5 fm.
Here, a larger R-matrix radius has to be used.
A range of 12 fm could indicate a polarization phenomenon.
More probably, it could just mean that the soft-sphere potential,
which generalizes the hard-sphere potential of R-matrix fits,
has no strong physical meaning either.
This seems confirmed by the fact that I could not find, up to now,
an $\ell$-independent soft-sphere potential describing the background in
several partial waves simultaneously.
A more promising way is to use a deep potential with bound and resonant states
corresponding to physical and Pauli-forbidden $^{16}$O states.
Such potentials can also be constructed by inversion
\cite{sparenberg:04a} and I plan to test their interest for
the hybrid model in the near future.


\end{document}